# Efficient spin-pumping and spin transport across epitaxial $Mn_3Sn(0001)$ noncollinear antiferromagnet/permalloy interfaces

Surya N. Panda[1,‡], Ning Mao[1], Nikolai Peshcherenko[1], Xiaolong Feng[1], Yang Zhang[2,3], Anastasios Markou[4], Claudia Felser[1], and Edouard Lesne[1,*]

[1]Max Planck Institute for Chemical Physics of Solids, 01187 Dresden, Germany

[2]Department of Physics and Astronomy, University of Tennessee, Knoxville, Tennessee 37996, USA

[3]Min H. Kao Department of Electrical Engineering and Computer Science, University of Tennessee, Knoxville, Tennessee 37996, USA

[4]Physics Department, University of Ioannina, 45110 Ioannina, Greece

## Abstract

The generation and control of spin currents are crucial for advancing next-generation spintronic technologies. These technologies depend on materials capable of efficiently sourcing and interconverting spin and charge currents, while overcoming some limitations associated with conventional ferromagnets and heavy metals. Kagome topological antiferromagnetic Weyl semimetals, such as $Mn_3Sn$, present unique advantages owing to their distinct magnetic order and significant Berry curvature-driven transport phenomena. In this study, we systematically investigate spin current generation and spin-to-charge conversion phenomena in epitaxial (0001)-oriented $Mn_3Sn$ thin films. Our findings reveal a moderate spin Hall angle of 0.9% and a nearly isotropic in-plane spin Hall conductivity of 44.4 ($\hbar$/e) $\Omega^{-1}.cm^{-1}$ at room temperature, originating from a combination of intrinsic and extrinsic contributions, as discussed in light of first-principle calculations. Furthermore, in $Mn_3Sn(0001)/Ni_{81}Fe_{19}$ heterostructures, we observe a high spin-mixing conductance of 28.52 $nm^{-2}$ and an interfacial spin-transparency of approximately 72%. Notably, we also find that the spin diffusion length in $Mn_3Sn(0001)$ epitaxial films exceeds 15 nm at room temperature. Our results highlight the potential, and limitations, of the topological Weyl noncollinear antiferromagnet $Mn_3Sn$ as an efficient material for spin transport and conversion in prospective spintronic applications.

‡ Current affiliation: Departments of Physics and Astronomy, Rice University, Houston, TX 77005, USA

* Correspondence and requests for materials should be addressed to: edouard.lesne@cpfs.mpg.de

## 1. Introduction

As magnetic storage and magnetic memory devices have become increasingly prevalent in the realm of computer electronics, a significant challenge persists: the requirement for high current densities to write magnetic bits. This challenge becomes particularly problematic as bit sizes decrease [1, 2]. In this context, spin-transfer and spin-orbit torque-based phenomena [3], particularly those driven by the spin Hall effect (SHE) [4], have garnered significant attention. These mechanisms facilitate magnetization reversal while alleviating the need for high current densities. Simultaneously, antiferromagnetic materials have emerged as a promising category of magnetically ordered systems for spin-orbitronic devices. Their advantageous properties—including minimal stray fields that allow for high bit density integration, terahertz dynamics (in contrast to gigahertz dynamics observed in ferromagnets), and resilience to external magnetic fields—render them highly appealing for the development of robust, high-density data storage devices [5, 6]. These characteristics are crucial for leveraging antiferromagnets (AFs) as sought-after spin-to-charge interconverters through SHE and inverse SHE. However, controlling the magnetic order in collinear AFs typically necessitates the application of large magnetic fields, often reaching several tens of Tesla. Furthermore, these materials typically do not exhibit electrical transport responses depending on the orientation of the Néel order parameter. This limitation restricts their capacity to generate substantial and reliable readout signals for magnetic memory applications [7].

A significant breakthrough has been achieved with the discovery of topological noncollinear AFs, such as $Mn_3X$ ($X$ = Sn, Ge, Ga) compounds. These materials demonstrate remarkable magneto-transport effects, including the intrinsic anomalous Hall effect (AHE), which arises from their unique magnetic structures and electronic topologies that encompass both Weyl nodes and nodal lines [8, 9]. In their noncollinear AF state, these compounds display (spin) Berry curvature hot-spots in reciprocal space. This characteristic not only contributes to the AHE but also facilitates intrinsic SHEs and substantial spin-orbit torques (SOT) [10–14]. Additionally, despite their vanishing net magnetization, these materials demonstrate sizeable anomalous Nernst effects [15, 16] driven by Berry curvature, as well as magneto-optical Kerr effects [17, 18], which are influenced by couplings to the octopular order parameter inherent in this class of noncollinear AFs. Notably, kagome noncollinear AFs possess chiral magnetic domains corresponding to different signs of the cluster octupole order parameter [14, 19, 20]. Consequently, the magneto-transport properties demonstrate reversed polarities across distinct domains. These intriguing characteristics position noncollinear AFs as key materials for the development of future spintronic devices.

$Mn_3Sn$ is a canonical noncollinear antiferromagnet that crystallizes in the hexagonal close-packed $D0_{19}$ structure (Fig. 1a) with space group $P6_3/mmc$ (No. 194). In this arrangement, Sn atoms occupy the centers of the hexagons formed by the Mn [21, 22]. Owing to geometric frustration in the kagome lattice, $Mn_3Sn$ adopts an inverse triangular spin configuration below its Néel temperature of approximately 420 K [23]. Additionally, it exhibits a weak ferromagnetic-like susceptibility in the plane, characterized by a small magnetic moment of approximately 0.28 $\mu_B$/f.u. [24] along an in-plane easy axis direction, owing to the overall uncompensated spins present in the basal plane of the inverse triangular AF order [25]. Despite extensive research on the magnetic and physical properties of $Mn_3Sn$, studies on magneto- and spin-transport phenomena have predominantly focused on bulk crystals of $Mn_3Sn$ and polycrystalline or textured $Mn_3Sn$ thin films. Consequently, the long-range-ordered epitaxial thin film counterparts remain largely unexplored.

In this study, we explore the spin-dependent transport properties of sputter-deposited (0001)–oriented $Mn_3Sn$ films, which feature a *c*-axis oriented out-of-plane and are grown heteroepitaxially on MgO(111) substrates. We conduct magneto-transport measurements from room temperature down to cryogenic temperature [26, 27]. Furthermore, we employ a combination of ferromagnetic resonance (FMR)-driven spin-pumping [28, 29] generation and inverse SHE (ISHE) [30] detection techniques. The thickness dependence of the Gilbert damping parameter indicates a notable combination of high interfacial spin transparency [31, 32] and substantial spin-mixing conductance at room temperature for $Mn_3Sn$/permalloy interfaces. Our findings reveal a sizeable spin Hall conductivity (SHC) in epitaxial $Mn_3Sn$ (0001) films, coupled with a nearly spin-transparent interface when in contact with permalloy. This underscores the potential of this canonical noncollinear AF as a candidate spin-to-charge converter for SOT-based device applications. The lack of anisotropy of the ISHE across two inequivalent crystallographic directions of hexagonal $Mn_3Sn$ suggests a competitive interplay between the intrinsic (I)SHE of topological origin driven by spin Berry curvature, and various extrinsic mechanisms, as discussed in this study.

## 2. Results

### 2.1. Structural properties of epitaxial $Mn_3Sn$(0001) films

To investigate the structural properties, crystallinity, heteroepitaxial relationships, and to determine the thickness and interfacial roughness of $Mn_3Sn$ heterostructures, we conducted various X-ray diffraction (XRD) measurements. The symmetric $2\theta$-$\omega$ XRD scans of a 40-nm-thick Mn3Sn film and a $Mn_3Sn$(40 nm)/Py(12 nm) heterostructure, both grown on a 3-nm-thick Ru underlayer on (111)–oriented cubic MgO single crystals are shown in Fig. 1b. The Ru underlayer and $Mn_3Sn$ film exclusively demonstrate the (0002) and (0004) reflections, indicative of the *c*-axis out-of-plane oriented hexagonal films. Additionally, a narrow full width at half maximum (FWHM) of 0.67° in the $\omega$-scan rocking curves around the $Mn_3Sn$ (0002) reflection for all film thicknesses can be observed in Fig. S1a of the supplementary information. This narrow FWHM suggests a high crystalline quality and low mosaicity. Furthermore, in the $Mn_3Sn$/Py bilayer, we observe that the Py film grows heteroepitaxially on the $Mn_3Sn$ film with a (111) texture, despite the growth of the latter occurring at room temperature. All samples are capped with an approximately 3-nm-thick Si layer (for detailed growth methods, see Section 5.1).

The epitaxial nature of the films is further corroborated by asymmetric XRD $\phi$-scans (Fig. 1c) of the $\{20\bar{2}1\}$ of the $Mn_3Sn$ film, $\{10\bar{1}1\}$ of the Ru underlayer, and the $\{202\}$ MgO substrate of the Bragg families of planes. Both the $Mn_3Sn$ film and Ru underlayer display six distinct reflections at 60° intervals, all aligning at the same azimuthal angle $\phi$. This six-fold symmetry validates the presence of hexagonal single-crystalline epilayers with well-defined in-plane orientation, which coincides with that of the three-fold in-plane symmetry of the cubic MgO(111) substrate. Consequently, we assert an epitaxial relationship between the substrate and epilayers, with the in-plane crystallographic directions aligning as follows: MgO(111)$[1\bar{1}0]$ ∥ Ru(0001)$[2\bar{1}\bar{1}0]$ ∥ $Mn_3Sn$(0001)$[2\bar{1}\bar{1}0]$, similar to the epitaxial relation for $Mn_3Sn$(0001) films grown on (111)-oriented $SrTiO_3$ substrates [24].

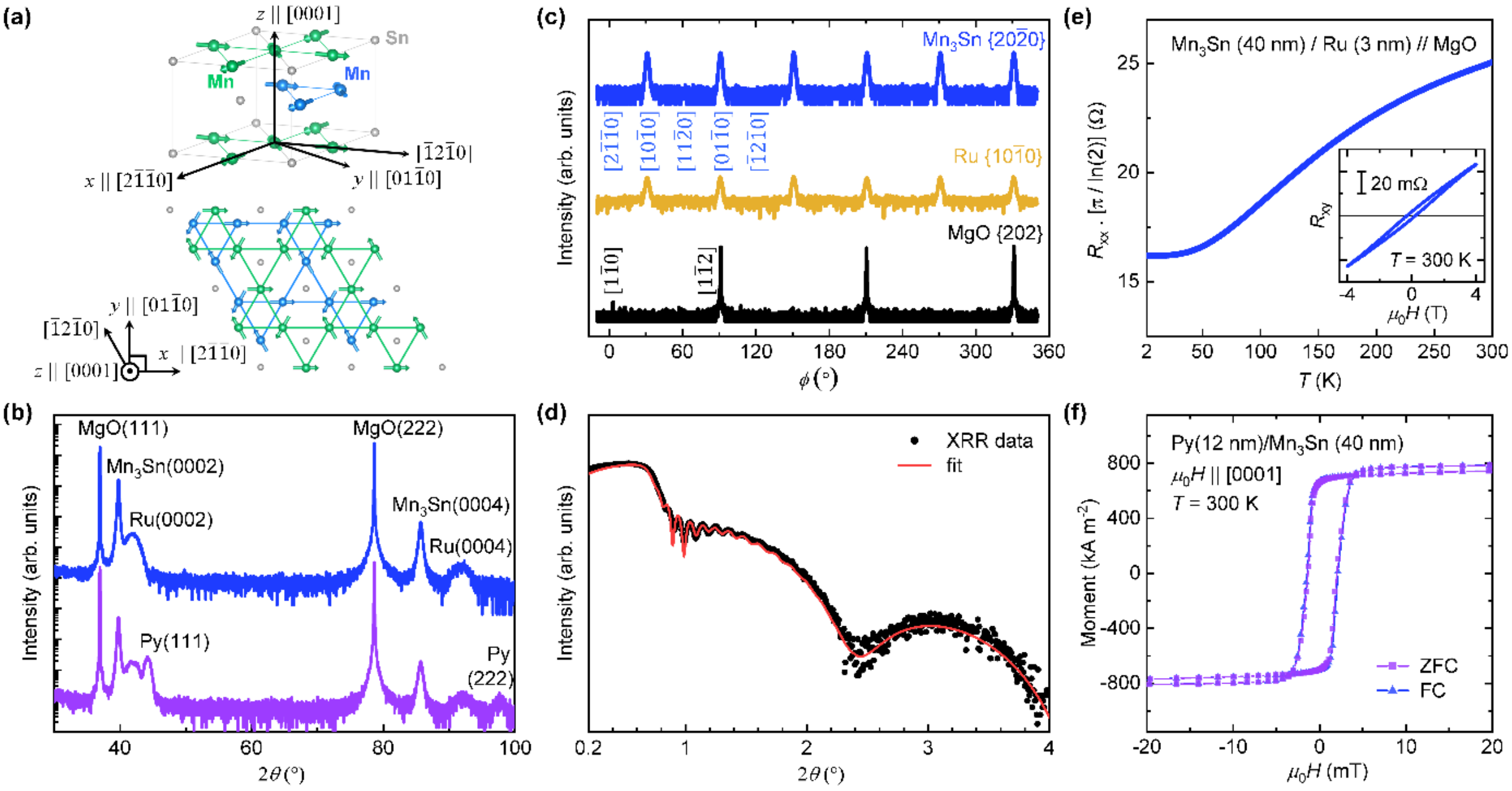

**Figure 1: Structural and charge transport properties of $Mn_3Sn$(0001) epitaxial thin films. a**, Crystal structure of hexagonal $Mn_3Sn$ (top) and projection on the (0001) plane (bottom). Mn atoms form a kagome-like plane and demonstrate an inverse triangular spin arrangement. **b**, XRD patterns of the $Mn_3Sn$(40 nm) thin films in the absence and presence of permalloy (Py) overlayer. **c**, Azimuthal $\phi$-scan patterns of the $Mn_3Sn\{20\bar{2}1\}$, Ru$\{10\bar{1}1\}$, and MgO$\{202\}$ Bragg families of planes. **d**, XRR pattern of the MgO//Ru(3 nm)/$Mn_3Sn$(40 nm)/$SiO_x$ heterostructure, where the red solid line represents the least-squares fit to the data. **e**, Longitudinal resistance ($R_{xx}$) as a function of temperature for a MgO//Ru(3 nm)/$Mn_3Sn$(40 nm)/$SiO_x$ sample. Inset: transverse (Hall) magnetoresistance ($R_{xy}$). **f**, Magnetic moment density at 300 K of a 12-nm-thick Py/40-nm-thick $Mn_3Sn$(0001) sample for field cooled and ZFC measurement protocols.

The X-ray reflectivity (XRR) for the MgO//Ru(3 nm)/$Mn_3Sn$(40 nm)/$SiO_x$(3 nm) sample is shown in Fig. 1d. The oscillatory intensity of the XRR indicates the conformal growth of well-defined $Mn_3Sn$ and Ru films with low roughness and sharp interfaces. The XRR data (symbols) is modeled using Parratt formalism (solid red line), allowing for the extraction of layer thickness, interface roughness, and electron density for each individual layer. The roughness values determined for the interfaces of MgO/Ru, Ru/$Mn_3Sn$, and $Mn_3Sn$/$SiO_x$ were 0.7, 0.9, and 1.2 nm, respectively. Additionally, the XRR fitting reveals the electron densities of the MgO, Ru, $Mn_3Sn$, and $SiO_x$ layers to be 3.8, 12.3, 6.5, and 2.8 g.cm$^{-3}$, respectively. The thickness of the $Mn_3Sn$ film was 39.5 nm, which will henceforth be referred to as a 40-nm-thick $Mn_3Sn$ sample. Atomic force microscopy (AFM) analysis, as shown in supplementary Fig. S1b, indicates an average root-mean-square topographical roughness of approximately 0.54 nm. Notably, the AFM topographical images do not reveal any evidence of long-range nonuniformity or dislocations within the $Mn_3Sn$ films, corroborating the findings from XRD and XRR characterizations.

The insights gained from both XRR and the (111)-textured nature of the permalloy (Py) film suggest the presence of a sharp, smooth interface between the $Mn_3Sn$ and Py layers. This interface not only mitigates the contributions from interfacial roughness to the exchange bias, as discussed below, but also potentially establishes a high-quality coherent interface that may enhance the efficiency of interfacial spin transport. Notably, the microstructure of materials is crucial for evaluating the (microscopic) origins of various phenomena, particularly in the context of spin currents, which are non-conserved quantities in condensed matter systems. With this in consideration, and following the structural characterizations previously presented, we conducted comprehensive electrical, magnetic, and spin-dependent transport characterizations of our $Mn_3Sn$(0001) epitaxial thin films.

## 2.2. Electrical transport properties of $Mn_3Sn$ (0001) films

We conducted temperature-dependent measurements of the longitudinal ($R_{xx}$) and transverse ($R_{xy}$) resistance of (111)–MgO//Ru/$Mn_3Sn$/$SiO_x$ samples with varying $Mn_3Sn$ thicknesses. The temperature ($T$-)dependent longitudinal resistance ($R_{xx}$) at zero field of a $SiO_x$-capped 40-nm-thick $Mn_3Sn$ film (with a 3-nm-thick Ru buffer layer) is shown in Fig. 1e. The $Mn_3Sn$ sample demonstrates metallic behavior, characterized by a continuous decrease in $R_{xx}$ ($T$) as the temperature is reduced from 320 to 2 K. The resistance of the $Mn_3Sn$ layer is calculated by utilizing a parallel resistor model (refer to supplementary Fig. S2 and supplementary information for details). Consequently, the resistivity of $Mn_3Sn$ is estimated to be 198.6 μΩ.cm at 300 K. The transverse magnetoresistance, $R_{xy}(\mu_0 H)$, shown in the inset of Fig. 1e, reveals a subtle spurious contribution from the anomalous Hall effect. This contribution demonstrates a slight hysteretic behavior, which may be attributed to the canting of magnetic moments away from the in-plane easy axis when subjected to sufficiently strong external out-of-plane magnetic fields. These observations align with the anisotropic AHE present in this noncollinear AF. In the Hall effect measurement configuration presented here, with the magnetic field along the [0001] crystallographic direction of $Mn_3Sn$, the Berry-curvature-driven intrinsic anomalous Hall conductivity is anticipated to diminish [8, 24].

## 2.3. Exchange bias in $Mn_3Sn$/Py bilayers

In AF/ferromagnet (FM) heterostructures, an exchange bias phenomenon can emerge at the interface owing to the exchange coupling between spins in the FM layer and uncompensated spins at the surface of the AF [33]. The presence of this exchange bias can significantly enhance the effective Gilbert damping of the FM, leading to a characteristic enhancement of the FMR linewidth. Additionally, this effect may introduce further asymmetry in the electrically detected inverse spin Hall voltage [34, 35]. Consequently, in bilayer systems demonstrating exchange bias, accurately quantifying the intrinsic magnetization dynamics—typically evaluated through FM resonance—and determining spin-to-charge interconversion efficiencies within the spin-pumping framework can become quite complex. To eliminate potential interferences from a putative exchange bias, we investigated $Mn_3Sn$(40 nm)/Py(12 nm) bilayers using superconducting quantum interference device (SQUID) magnetometry. Our measurement protocol begins with the demagnetization of the heterostructure at 400 K in a strong in-plane oscillatory magnetic field. This was followed by a cooldown of the sample to the measurement temperature, either in a zero external magnetic field or under a 1 T in-plane field. The resulting zero-field cooled (ZFC) and field-cooled in-plane hysteresis loops are obtained at 300 K (as shown in Fig. 1f). These results indicate the absence of a detectable exchange bias at room temperature, aligning with a previous report by Markou *et al.* in similar epitaxial $Mn_3Sn$(0001)/Py thin film heterostructures [36]. Therefore, the FMR-driven spin-pumping measurements conducted at 300 K are free from spurious exchange bias contributions.

## 2.4. FMR-driven spin-pumping measurements

Having established that the $Mn_3Sn$ films demonstrate a unique noncollinear AF spin structure and a zero exchange bias when in contact with Py at room temperature, we now focus on the generation of pure spin currents injected from Py into $Mn_3Sn$ through FMR spin-pumping, and their conversion into a charge current through the ISHE in $Mn_3Sn$. Pure spin currents, which involve only the flow of spin angular momentum with no charge movement, are essential for the development of energy-efficient spintronic devices. They effectively mitigate the limitations associated with charge-based devices, such as Joule heating and stray Oersted fields [37]. Among various methods for generating spin currents, the spin-pumping mechanism emerges as an effective technique. This primarily results from its ability to circumvent the challenges associated with impedance mismatch, enabling the generation of pure spin currents across macroscopically broad areas without the necessity for lithographically patterned samples. At the FMR of the FM layer, the transfer of spin angular momentum to an adjacent metallic layer provides new dissipation channels for the out-of-equilibrium magnetization, resulting in a pure diffusive spin current. Consequently, this process increases the effective Gilbert damping parameter of the FM layer. The efficiency of this spin-pumping mechanism is characterized by the spin-mixing conductance ($g_{\uparrow\downarrow}$), which, along with the interfacial spin-transparency ($\eta$), governs the magnitude of the injected pure spin current at the interface. We progressively introduce these relevant concepts in the following sections.

In this study, we recorded the magnetization dynamics using a NanOsc FMR setup (refer to Methods Section 5.4). A characteristic subset of FMR spectra for the $Mn_3Sn$(40 nm)/Py(12 nm) sample, recorded between 4 and 20 GHz excitation frequencies, is shown in Fig. 2a. The collected FMR spectra represent the field derivative of the imaginary part ($\chi$") of the dynamic magnetic susceptibility ($\chi$) as a function of the externally applied DC magnetic field ($\mu_0 H$). For each FMR spectrum, the linewidth ($\Delta H$) and resonance field ($H_{res}$) can be extracted from the data using the following formula [38]:

$$\frac{d\chi"}{dH}(H) = K_{\text{abs}} \frac{4\Delta H(H-H_{\text{res}})}{[4(H-H_{\text{res}})^2+(\Delta H)^2]^2} - K_{\text{dis}} \frac{(\Delta H)^2 - 4(H-H_{\text{res}})^2}{[4(H-H_{\text{res}})^2+(\Delta H)^2]^2}, \quad (1)$$

where $K_{abs}$ and $K_{dis}$ represent coefficients to the field-antisymmetric and field-symmetric components, respectively, linked to absorptive and dispersive contributions to $\chi_m$. A field-linear background and field-independent vertical offset were employed for the fitting procedure of $\mathrm{d}\chi"/\mathrm{d}H(H)$. However, this aspect has been omitted in Eq. 1 for simplicity. Notably, $\Delta H$ represents the full-width at half-maximum (FWHM) of the Lorentzian functions, which is related to the peak-to-peak linewidth of the FMR spectra as: $\Delta H_{pp} = \Delta H_{\text{FWHM}}/\sqrt{3}$. The corresponding extracted resonance field ($\mu_0 H_{res}$) dependence on the resonance frequency, $f_{res}$, is shown in Fig. 2b for the $Mn_3Sn$(40 nm)/Py(12 nm) heterostructure, alongside a reference 12-nm-thick Py film (capped with $SiO_x$). Further, the effective saturation magnetization value ($M_{\text{eff}}$) and the anisotropy field ($H_k$) can be determined by fitting the dispersion relation $f_{res}(H_{res})$ using Kittel's formula [39]:

$$f_{\text{res}} = \frac{\gamma}{2\pi}\sqrt{(H_{\text{res}} + H_k)(H_{\text{res}} + H_k + 4\pi M_{\text{eff}})}\,, \quad (2)$$

where $\gamma = \frac{g_{\text{Py}}\mu_{\text{B}}}{\hbar}$ represents the gyromagnetic ratio, $g_{\text{Py}}$ = 2.11 represents the electron $g$-factor of permalloy [40], $\mu_B$ represents the Bohr magneton, $\hbar$ represents the reduced Planck constant, and $\mu_0$ represents the vacuum permeability. Following this approach, we obtained consistent values of the effective saturation magnetization $M_{eff}$ = 756.1 kA.m$^{-1}$ for the bilayer system in the presence of $Mn_3Sn$, and 738.5 kA.m$^{-1}$ for solely the Py. The value of $|\mu_0 H_k|$ obtained from applying the Kittel formula is of the order of 1 mT for our Py and $Mn_3Sn$/Py samples.

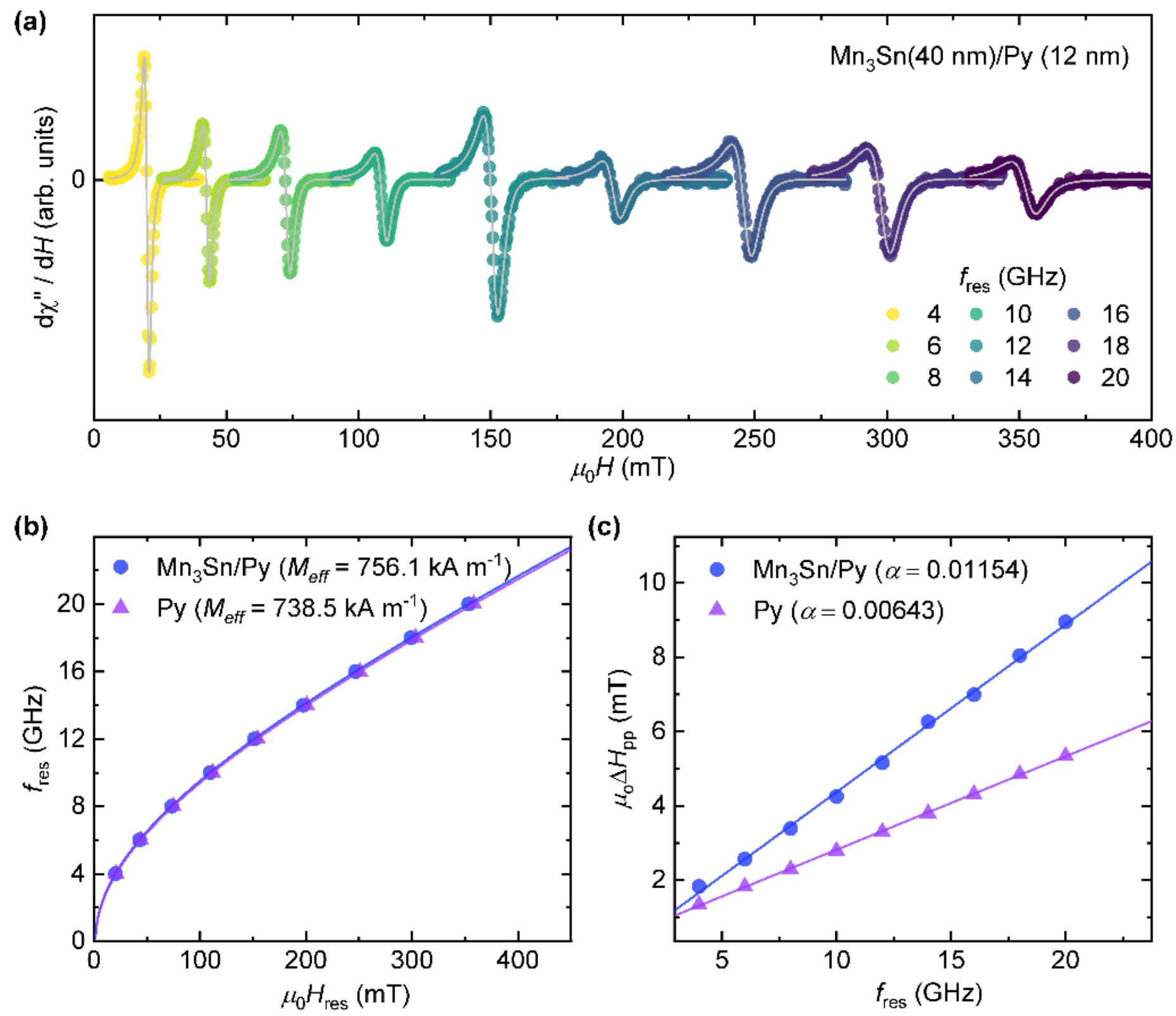

**Figure 2: Broadband FMR measurements. a**, FMR spectra of the $Mn_3Sn$(40 nm)/Py(12 nm) sample at different excitation frequencies. The symbols represent experimental data and solid lines indicates fits using Eq. 1. **b**, Resonance frequency versus magnetic field for $Mn_3Sn$ (40 nm)/Py(12 nm) and Py(12 nm), with solid lines fitted using Kittel's formula, Eq. 2. **c**, Peak-to-peak linewidth versus resonance frequency, with solid lines fitted using Eq. 3 to determine the effective Gilbert damping parameter ($\alpha$).

Subsequently, we experimentally estimate the Gilbert damping parameter ($\alpha$) of Py by fitting the peak-to-peak FMR linewidth $\Delta H_{\mathrm{pp}}$ as a function of $f_{\mathrm{res}}$, as follows [41]:

$$\Delta H_{\mathrm{pp}} = \frac{4\pi}{\gamma\sqrt{3}}\alpha f_{\mathrm{res}} + \Delta H_0 \,. \tag{3}$$

$\Delta H_0$ represents a frequency-independent inhomogeneous linewidth broadening, which is associated with the magnetic inhomogeneity of the heterostructure. The gyromagnetic ratio was derived from the Kittel formula. Consequently, the slope in Eq. 3 is determined by the magnitude of the effective Gilbert damping parameter $\alpha$. The $\mu_0\Delta H_{\mathrm{pp}}$ versus $f_{\mathrm{res}}$ for the $Mn_3Sn$ (40 nm)/Py (12 nm) and Py (12 nm) samples are shown in Fig. 2c. The linear behavior of $\Delta H_{\mathrm{pp}}(f_{\mathrm{res}})$ suggests a high degree of homogeneity in our samples. Notably, $\mu_0\Delta H_0$ accepts values below 0.4 mT, both in the presence and absence of the $Mn_3Sn$ layer. The determined value of $\alpha$ for the $Mn_3Sn$ (40 nm)/Py(12 nm) was $0.01154 \pm 0.00018$, significantly higher than $0.00643 \pm 0.00004$ observed for the Py(12 nm) reference sample. This disparity in the Gilbert damping parameter values strongly suggests the presence of an efficient spin-pumping effect [42]. The increase in the Gilbert damping parameter can be interpreted as the result of the flow of spin angular momentum density—essentially a diffusive spin current— across the interface, which exerts an additional damping-like torque on the magnetization of the Py layer. However, other spurious contributions may also increase $\alpha$, as discussed in more detail in the subsequent sections. Furthermore, no significant variations in the magnitudes of $\alpha$ or $M_{eff}$ were observed when measured along the $[2\bar{1}\bar{1}0]$ and $[01\bar{1}0]$ in-plane crystallographic directions of $Mn_3Sn$ (see supplementary Fig. S3). This finding suggests that in-plane anisotropy does not significantly influence spin transport in $Mn_3Sn$/Py heterostructures.

## 2.5. Inverse spin Hall effect measurements

In FMR-driven spin-pumping experiments, the measurement of the ISHE is widely recognized for detecting the conversion of spin current into charge current. Within the framework of the SHE, materials demonstrating substantial spin-orbit interaction cause electrons with opposite spins to be deflected in opposite directions. This results in a pure spin current ($I_s$) that is transverse to that of the charge current ($I_c$). The phenomenon known as the ISHE originates from the Onsager reciprocal relations, facilitating the conversion of spin currents into transverse charge currents. Notably, the intrinsic source term for both effects is the spin Berry curvature [43]. A commonly employed figure of merit for quantifying the efficiency of spin-charge interconversion is the spin Hall angle (SHA; $\theta_{SH}$), defined as: $\theta_{SH}$ = $\left({}^{I_s}/_{I_c}\right)$. The determination of this parameter will be addressed in the following section.

We measure the ISHE in conjunction with FMR detection by evaluating the DC voltage drop (related to $I_c$ in an open electrical circuit) across the metallic $Mn_3Sn$/Py heterostructure, as shown in Fig. 3a (refer to Methods for further details). The detected overall voltage drop ($\Delta V$) versus $\mu_0 H$ is fitted with a combination of Lorentzian and anti-Lorentzian functions [44, 45]:

$$\Delta V = V_{\mathrm{sym}}\left[\frac{(\Delta H)^2}{(H-H_{\mathrm{res}})^2+(\Delta H)^2}\right] + V_{\mathrm{as}}\left[\frac{\Delta H(H-H_{\mathrm{res}})}{(H-H_{\mathrm{res}})^2+(\Delta H)^2}\right] . \tag{4}$$

The field-symmetric amplitude $V_{\mathrm{sym}}$ primarily represents the ISHE contribution, effectively excluding other spurious rectification effects (*e.g.*, anisotropic magnetoresistance or thermal effects). Therefore, $V_{ISHE} \cong V_{sym}$. The field-antisymmetric amplitude $V_{\mathrm{as}}$ is a product of both the anomalous and planar Hall effects. For simplicity, Eq. 4 does not include a field-linear term and an offset voltage; however, these factors are accounted for in fitting of the experimental data. The acquired voltage for both positive and negative magnetic field polarities of the $Mn_3Sn$ (40 nm)/Py (12 nm) sample are shown in Fig. 3b. The fitting and decomposition of $\Delta V$, in accordance with Eq. 4, facilitate a direct estimation of the magnitude of $V_{\mathrm{ISHE}}$ around the FMR field of Py, thereby validating that the observed DC voltage results from spin-pumping/ISHE processes. Notably, the sign of $V_{ISHE}$ reverses upon inverting the magnetic field polarity, which aligns with the expectations of the ISHE framework that as the spin-polarization ($\vec{s}$) axis is also reversed, the relationship $\vec{J_c} \propto (\vec{J_s} \times \vec{s}\,)$ holds.

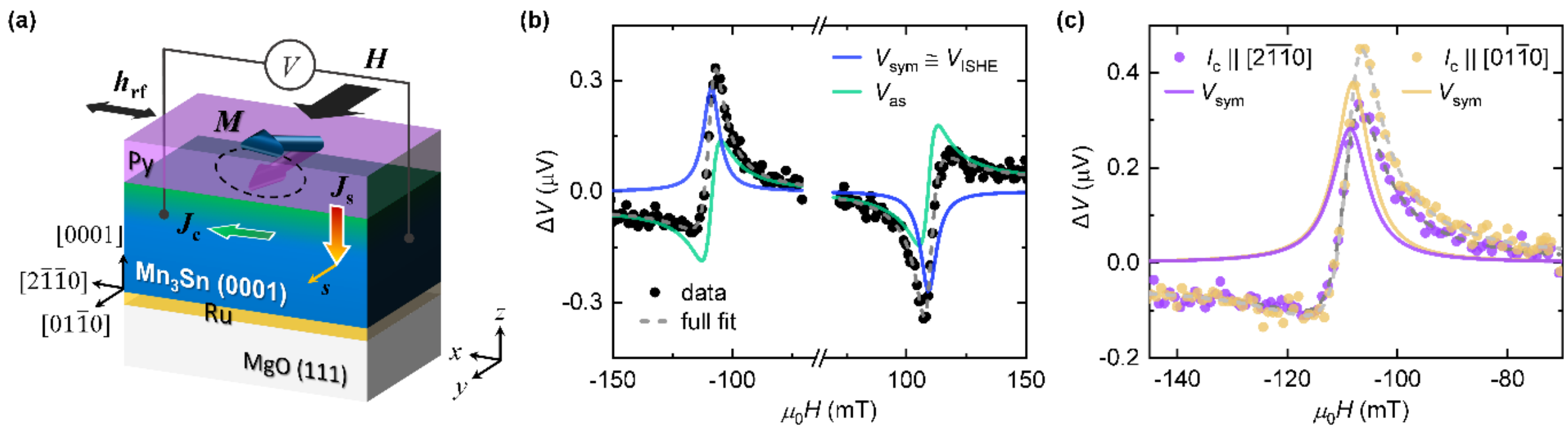

**Figure 3: FMR-driven spin-pumping and ISHE measurements. a**, Schematic of the sample structure for FMR-driven spin-pumping and ISHE measurements. **b**, Voltage drop detected in FMR-spin-pumping experiments for the $Mn_3Sn$(40 nm)/Py(12 nm) sample at 10 GHz and 300 K. Decompensation of the measured voltage in terms of its field-symmetric ($V_{sym} \cong V_{ISHE}$) and antisymmetric ($V_{as}$) parts, displayed for positive and negative magnetic field polarities. The dashed gray lines represent the fit using Eq. 4, whereas the solid blue and green lines represent the symmetric and antisymmetric contributions, respectively. **c**, Detected voltage drops (scattered points) along the $[01\bar{1}0]$ and $[2\bar{1}\bar{1}0]$ crystallographic directions of $Mn_3Sn$, highlighting the quasi-isotropic magnitude and similar sign of $V_{sym}$ (solid lines).

Notably, on the one hand, the noncollinear AF structure of $Mn_3Sn$ demonstrated in-plane uniaxial magnetic anisotropy owing to the small uncompensated moment along the $[2\bar{1}\bar{1}0]$. This anisotropy is reflected in the slightly anisotropic magneto-transport properties depending on the direction of the in-plane applied field [24]. On the other hand, theoretical and symmetry analyses suggest that the spin-dependent intrinsic transport responses, characterized by the spin conductivity tensor, are expected to demonstrate significant anisotropic behaviors in hexagonal noncollinear AFs [10, 11]. This topic will be explored in greater detail in Sections 2.7 and 3. However, when measured along the $[2\bar{1}\bar{1}0]$ and $[01\bar{1}0]$ orthogonal crystallographic axis of $Mn_3Sn$ films, the sign of $V_{ISHE}$ remains unchanged, and the magnitude is comparable (within 25%), as shown in Fig. 3c, at variance with predictions given in Ref. [10]. The observed quasi-isotropic response suggests that, despite the crystalline and epitaxial nature of the $Mn_3Sn$(0001) films under investigation, the spin-to-charge conversion mechanism through ISHE in our heterostructures is not dominated by the intrinsic spin Berry curvature mechanism. While this experimentally observed nearly isotropic behavior has profound implications in terms of the potential sources of the ISHE in $Mn_3Sn$ epitaxial films (developed in Section 2.7), the thickness-dependent spin transport results hereinafter are presented for a fixed measurement geometry, with the charge current propagation direction along the $[2\bar{1}\bar{1}0]$ crystallographic axis of $Mn_3Sn$.

## 2.6. Interfacial spin transport at $Mn_3Sn$(0001)/Py interfaces

Before delving into the efficiency of the spin-to-charge conversion in $Mn_3Sn$, we examine the processes involved in interfacial spin transport, as these processes ultimately determine the magnitude of the detected signals. The transfer of spin angular momentum across the $Mn_3Sn$/Py interface, without spin backflow, is characterized by the intrinsic spin-mixing conductance, denoted as $g_{\uparrow\downarrow}$. This intrinsic $g_{\uparrow\downarrow}$ quantifies the conductance properties of spin channels at the interface when the thickness of the spin sink layer (here: $Mn_3Sn$) significantly exceeds the spin diffusion length ($\lambda_{\mathrm{sd}}$) within $Mn_3Sn$. In the presence of spin backflow, particularly when the thickness of $Mn_3Sn$ is comparable to or smaller than or of the order of $\lambda_{\mathrm{sd}}$, interfacial spin transport is instead governed by an effective spin-mixing conductance, denoted $g_{\uparrow\downarrow}^{\mathrm{eff}}$, which depends on the overall material properties, interface considered, and thickness of the spin-to-charge converter [42, 46, 47].

The value of $g_{\uparrow\downarrow}$ can be derived from the thickness dependence of the Gilbert damping (Fig. 4a). This analysis employs a simplified ideal spin sink approximation, represented as follows [48]:

$$\alpha = \alpha_0 + \frac{g_{\mathrm{Py}}\mu_{\mathrm{B}}g_{\uparrow\downarrow}\left(1-e^{-\frac{2t}{\lambda_{\mathrm{sd}}}}\right)}{4\pi M_{\mathrm{eff}}d_{\mathrm{Py}}}, \tag{5}$$

where $\alpha_0$ represents the intrinsic Gilbert damping constant of Py; $t$ and $\lambda_{\mathrm{sd}}$ represent the $Mn_3Sn$ layer thickness and the spin diffusion length in $Mn_3Sn$, respectively; and $d_{\mathrm{Py}}$ = 12 nm represents the thickness of Py. The model assumes that spin angular momentum is effectively transferred to the spin sink layer, where it experiences spin decoherence. This implies that the back-flow of spin current owing to backscattering is minimal, and the enhancement of Gilbert damping is predominantly influenced by the pure spin current resulting from an efficient spin-pumping process. The $Mn_3Sn$ thickness-dependent modulation of $\alpha$ in $Mn_3Sn$/Py heterostructures, demonstrating the characteristic saturation behavior predicted by the model corresponding to Eq. 5, is shown in Fig. 4a. This behavior culminates in an asymptotic saturated value of 0.01132. Through the fitting procedure, we obtain $g_{\uparrow\downarrow}$ = 28.52 nm$^{-2}$ for the $Mn_3Sn$(0001)/Py(111) interface, whereas $\lambda_{\mathrm{sd}}$ in $Mn_3Sn$ reached 25.3 nm, which is relatively large compared with common materials utilized for spintronics [49]. We will further critically evaluate the estimated magnitude of $\lambda_{\mathrm{sd}}$ in relation to the thickness-dependence of the generated charge current.

We now focus on estimating the SHA and the associated SHC. The generated charge current $I_c$ from the ISHE process is determined by the ratio of the detected symmetric voltage component $V_{\mathrm{sym}}$ to the sheet resistance of $Mn_3Sn$, $I_{\mathrm{c}} = (V_{\mathrm{sym}}t)/\rho_{\mathrm{Mn_3Sn}}$, with $t$ representing the thickness of the $Mn_3Sn$ layer. The $Mn_3Sn$ thickness-dependent variation in $I_{\mathrm{c}}$ is shown in Fig. 4b. This curve indicates a non-monotonic increase that eventually saturates at high thicknesses. This behavior for $I_{\mathrm{c}}(t)$, which is linked to the finite spin diffusion length in $Mn_3Sn$, can be employed to determine the effective spin Hall angle ($\theta_{\mathrm{SH}}^{\mathrm{eff}}$) of the $Mn_3Sn$ layer, as these parameters are related by the following equation [50, 51]:

$$I_{\mathrm{c}}/w = \theta_{\mathrm{SH}}^{\mathrm{eff}}\lambda_{\mathrm{sd}}\tanh\left[\frac{t}{2\lambda_{\mathrm{sd}}}\right]I_{\mathrm{s}}\ , \tag{6}$$

where $I_{\mathrm{s}}$, the spin current generated by spin-pumping, is expressed as follows:

$$I_s = \left(\frac{g_{\uparrow\downarrow}\hbar}{8\pi}\right)\left(\frac{\mu_0 h_{rf}\gamma}{\alpha}\right)^2 \left[\frac{4\pi M_{eff}\gamma+\sqrt{(4\pi M_{eff}\gamma)^2+(4\pi f_{res})^2}}{(4\pi M_{eff}\gamma)^2+(4\pi f_{res})^2}\right]\left(\frac{2e}{\hbar}\right), \tag{7}$$

where $w$ represents the width of the sample atop the coplanar waveguide and $h_{rf}$ represents the magnitude of the RF magnetic field (calibrated to be 61 μT in the frequency range utilized here). However, the other quantities have been determined experimentally. The dependence of $I_c$ on the thickness of $Mn_3Sn$ layers is accurately represented in Eq. 6, as indicated by the solid line in Fig. 4b. From this fitting procedure, we determine the effective SHA ($\theta_{SH}^{eff}$) of $Mn_3Sn$ to be 0.63 ± 0.07%, whereas the spin diffusion length $\lambda_{sd}$ = 15.2 ± 0.2 nm. The discrepancy in the estimated values of $\lambda_{sd}$ , particularly compared with those derived from the thickness-dependence of the Gilbert damping, presents a complex challenge that cannot be easily resolved.

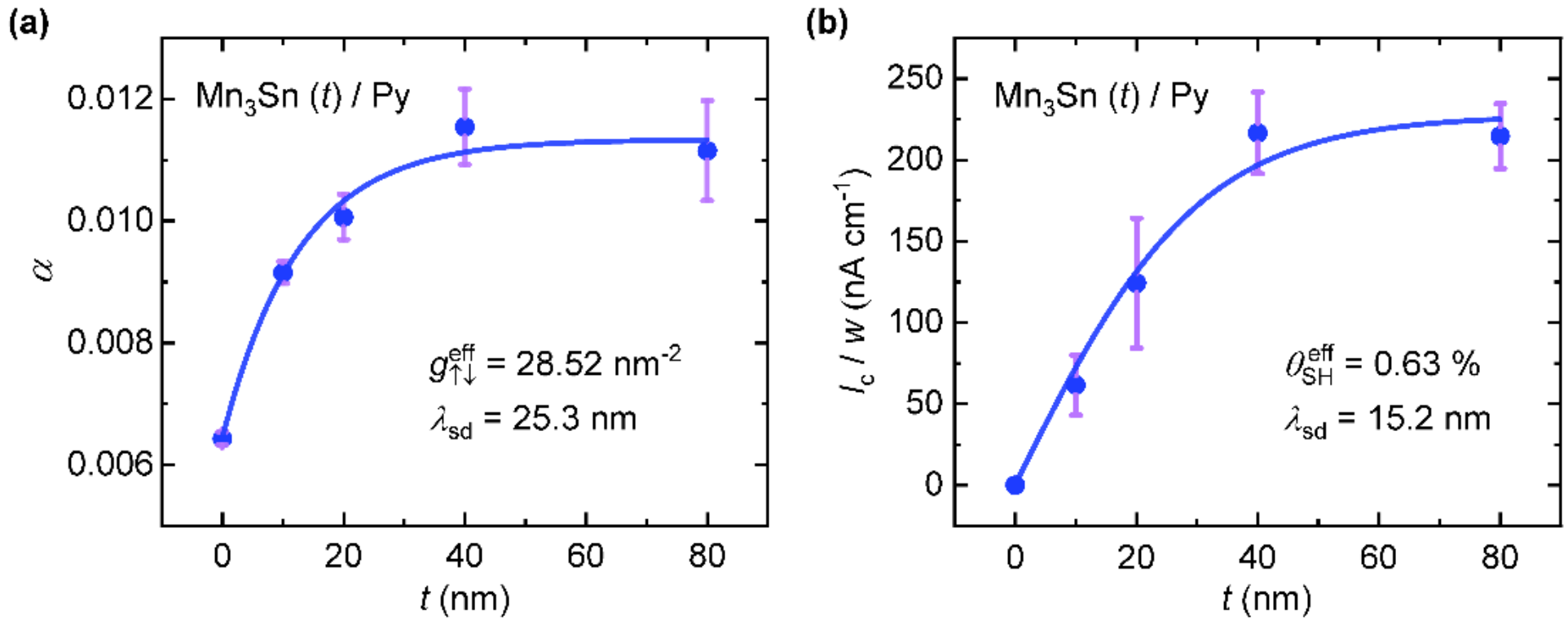

**Figure 4: $Mn_3Sn$ thickness-dependent spin-pumping and ISHE results. a**, $Mn_3Sn$ thickness-dependent modulation of the Gilbert damping parameter ($\alpha$); the solid line represents the fit using Eq. 5. **b,** $Mn_3Sn$ thickness-dependent modulation of ISHE-induced charge current ($I_c$) at 300 K. The solid line corresponds to the fit using Eq. 6, and returns the estimate of the magnitude of the effective SHA.

The effective thickness (or volume) region, which enhances the Gilbert damping, may differ from the region involved in spin-to-charge conversion generating the resultant charge current. This divergence may stem from various mechanisms that can lead to similar observed differences in $\lambda_{sd}$ and SHAs. Previous studies extensively discuss these mechanisms, which include ballistic [49] versus diffusive regime approximations [50, 51], nonequilibrium (inverse) magnetic proximity effects under simultaneous SHE and ISHE [52], and inclusion or exclusion of additional interfacial spin decoherence or spin-flip scattering processes. Furthermore, the broader concepts of "spin memory loss" [53, 54] versus "interfacial spin transparency" [31, 32] have also been explored. Disentangling these contributions—whether they are microscopic or phenomenological—remains a significant experimental challenge, even in relatively straightforward FM/nonmagnetic interfaces. This complexity is compounded by the presence of other electric field-driven sources of spin and charge currents that may contribute beyond the (I)SHE processes [55–57].

In general, the generation of pure spin currents through spin pumping does not guarantee that all spins accumulating at the $Mn_3Sn$/Py interface effectively diffuse into the $Mn_3Sn$ layer and undergo spin-to-charge conversion. In a more realistic scenario, a portion of this spin accumulation may be lost owing to factors, such as interfacial disorder, surface roughness, or intrinsic electronic band misalignments at the interface. The concept of spin-transparency, denoted as $\eta$, encompasses these effects and quantifies the overall probability of spins being reflected or transmitted at the interface. In a diffusive spin-transport model, $\eta$ can be expressed as a function of $g_{\uparrow\downarrow}$ and $\lambda_{\mathrm{sd}}$ as follows [31, 32]:

$$\eta = \frac{g_{\uparrow\downarrow} \tanh\left(\frac{t}{2\lambda_{sd}}\right)}{g_{\uparrow\downarrow}\coth\left(\frac{t}{\lambda_{sd}}\right)+\frac{h}{2\lambda_{sd}e^{2}\rho}} , \quad (8)$$

where $\rho$ represents the electrical resistivity of the $Mn_3Sn$ layer, and $t$ its thickness, $h$ represents Planck's constant, and $e$ represents the elementary charge. Utilizing the experimentally obtained values for $g_{\uparrow\downarrow}$ and $15.2 \leq \lambda_{\mathrm{sd}} \leq 25.3$ nm (Fig. 4a and Eq. 5), we obtain $\eta = 72 \pm 12\%$ for the 40-nm-thick $Mn_3Sn$ film. This estimate assumes ideal spin-sink conditions, where $t \gg \lambda_{sd}$, and corresponds to the regime in which Gilbert damping modulation reaches saturation. In retrospect, the notably high value of the spin transparency observed in this study may be attributed to the superior quality of the interface in the epitaxial $Mn_3Sn$(0001)/Py(111) heterostructures and provides a tentative guide for enhancing spintronics devices whose efficiency inherently relies on interface quality.

By leveraging the determined value of $\eta$, we estimated the corrected SHA of $Mn_3Sn$ to be $\theta_{\mathrm{SH}} = \left(\theta_{\mathrm{SH}}^{\mathrm{eff}}/\eta\right) = 0.88 \pm 0.17\%$. This value is notably lower than the previously reported SHA for polycrystalline $Mn_3Sn$ films ($\theta_{\mathrm{SH}} = 5.3 \pm 2.4\%$). However, these polycrystalline films demonstrate a significantly shorter spin diffusion length ($\lambda_{\mathrm{sd}} \approx 0.75$ nm), and a considerably higher electrical resistivity—over five times greater than that of our epitaxial films [58]. In such disordered films, substantial contributions can be anticipated from mechanisms, such as spin-flip and side-jump scattering, owing to grain boundaries and anti-site disorder acting as effective impurities, which add to the intrinsic spin Berry curvature-driven contribution. In hexagonal polycrystalline films with low symmetry, as opposed to cubic systems, the (I)SHEs are influenced by a combination of multiple intrinsic SHC tensor elements, resulting from the simultaneous response from multiple inequivalent crystallographic orientations. In our $Mn_3Sn$(0001) epitaxial films, we estimated the magnitude of the SHC as $\sigma_{\mathrm{SH}} = \frac{\theta_{\mathrm{SH}}}{\rho_{\mathrm{Mn_3Sn}}}\left(\frac{\hbar}{e}\right) \approx 44.4$ ($\hbar$/e) $\Omega^{-1}$.cm$^{-1}$. All the spin-dependent parameters and figures of merit obtained from this study are listed in Table S1 of the supplementary information, and compared to a number of studies of spin-charge interconversion experiments in $Mn_3Sn$. Our findings demonstrate a favorable comparison with other studies on spin-dependent transport in $Mn_3Sn$ and underscore the overall potential and competitiveness of epitaxial $Mn_3Sn$ thin films for future spintronics devices. We note that the strong scattering of values reported in the literature for *e.g.*, $\theta_{\mathrm{SH}}$ and $\sigma_{\mathrm{SH}}$ magnitudes (see supplementary Table S1) can tentatively be attributed in some cases to $Mn_3Sn$ (ultra-thin) films of uneven crystallographic ordering, microstructure, and overall greater disorder, which do not easily compare with our long-range ordered epitaxial $Mn_3Sn$(0001) films.

## 2.7. Origins of ISHE in $Mn_3Sn$(0001) films: insights from first-principle calculations

To elucidate the mechanisms underlying spin-to-charge interconversion mechanisms in $Mn_3Sn$, we conducted density functional theory (DFT) calculations utilizing the Vienna *ab initio* simulation package for $Mn_3Sn$ in the hexagonal space group $P6_3/mmc$ (No. 194). We subsequently derived an effective tight-binding Hamiltonian to compute the spin-dependent transport coefficients through the Kubo linear response approach (Methods Section 5.5). The origin of the intrinsic SHC of hexagonal noncollinear $Mn_3X$ ($X$ = Sn, Ge, Ga) systems has been thoroughly predicted and analyzed by Zhang *et al.* [10]. The SHC tensor, denoted as $\sigma_{\alpha\beta}^{\gamma}$, is a rank-3 tensor. It characterizes the relationship between the electric field (or corresponding charge current induced by the ISHE) along $x$-direction and the spin current polarized along $y$-direction, which flows along the out-of-plane $z$-direction. This relationship can be expressed as: $J_z^y = \sigma_{zx}^y E_x$ (refer to Methods Section 5.5). Specifically, the intrinsic SHC tensor, which arises from spin Berry curvature, demonstrates even symmetry under time-reversal (TR) operations. Owing to the additional symmetry constraints inherent to hexagonal $Mn_3Sn$, many elements of the SHC tensor are compelled to vanish, resulting in only six independent non-zero elements for the TR-even SHC tensor, as presented in Table 1.

**Table 1:** Symmetry-based spin Berry curvature-driven intrinsic (TR even) and disorder-driven extrinsic (TR-odd) spin Hall conductivity tensors of a noncollinear AF hexagonal $Mn_3Sn$. The $z$-direction corresponds to the out-of-plane [0001] direction of $Mn_3Sn$, whereas the orthogonal in-plane $x$- and $y$-directions corresponds to the $[2\bar{1}\bar{1}0]$ and $[01\bar{1}0]$ crystallographic axes of $Mn_3Sn$, respectively. For clarity: the superscript index corresponds to the spin-polarization axis, whereas the first and second subscript indices represent the spin and charge current flow directions, respectively.

| SHC tensor | $\sigma^x$ | $\sigma^y$ | $\sigma^z$ |
|---|---|---|---|
| Intrinsic TR-even | $\begin{pmatrix} 0 & 0 & 0 \\ 0 & 0 & \sigma_{yz}^x \\ 0 & \sigma_{zy}^x & 0 \end{pmatrix}$ | $\begin{pmatrix} 0 & 0 & \sigma_{xz}^y \\ 0 & 0 & 0 \\ \sigma_{zx}^y & 0 & 0 \end{pmatrix}$ | $\begin{pmatrix} 0 & \sigma_{xy}^z & 0 \\ \sigma_{yx}^z & 0 & 0 \\ 0 & 0 & 0 \end{pmatrix}$ |
| Extrinsic TR-odd | $\begin{pmatrix} 0 & \sigma_{xy}^x & 0 \\ \sigma_{yx}^x & 0 & 0 \\ 0 & 0 & 0 \end{pmatrix}$ | $\begin{pmatrix} \sigma_{xx}^y & 0 & 0 \\ 0 & \sigma_{yy}^y & 0 \\ 0 & 0 & \sigma_{zz}^y \end{pmatrix}$ | $\begin{pmatrix} 0 & 0 & 0 \\ 0 & 0 & \sigma_{yz}^z \\ 0 & \sigma_{zy}^z & 0 \end{pmatrix}$ |

We remark that the $\sigma^z$ tensor elements were inaccessible within our spin-pumping geometry, which constrains the spin-polarization axis remained in-plane. Consequently, only two tensor elements could be experimentally measured through the combined spin-pumping and ISHE: $\sigma^x_{zy}$ and $\sigma^y_{zx}$. We will discuss the TR-odd SHC, as listed in Table 1, in greater detail. The magnitude of this SH conductivity is influenced by disorder-related factors, yet it fundamentally originates from the specific noncollinear inverse triangular AF order [11].

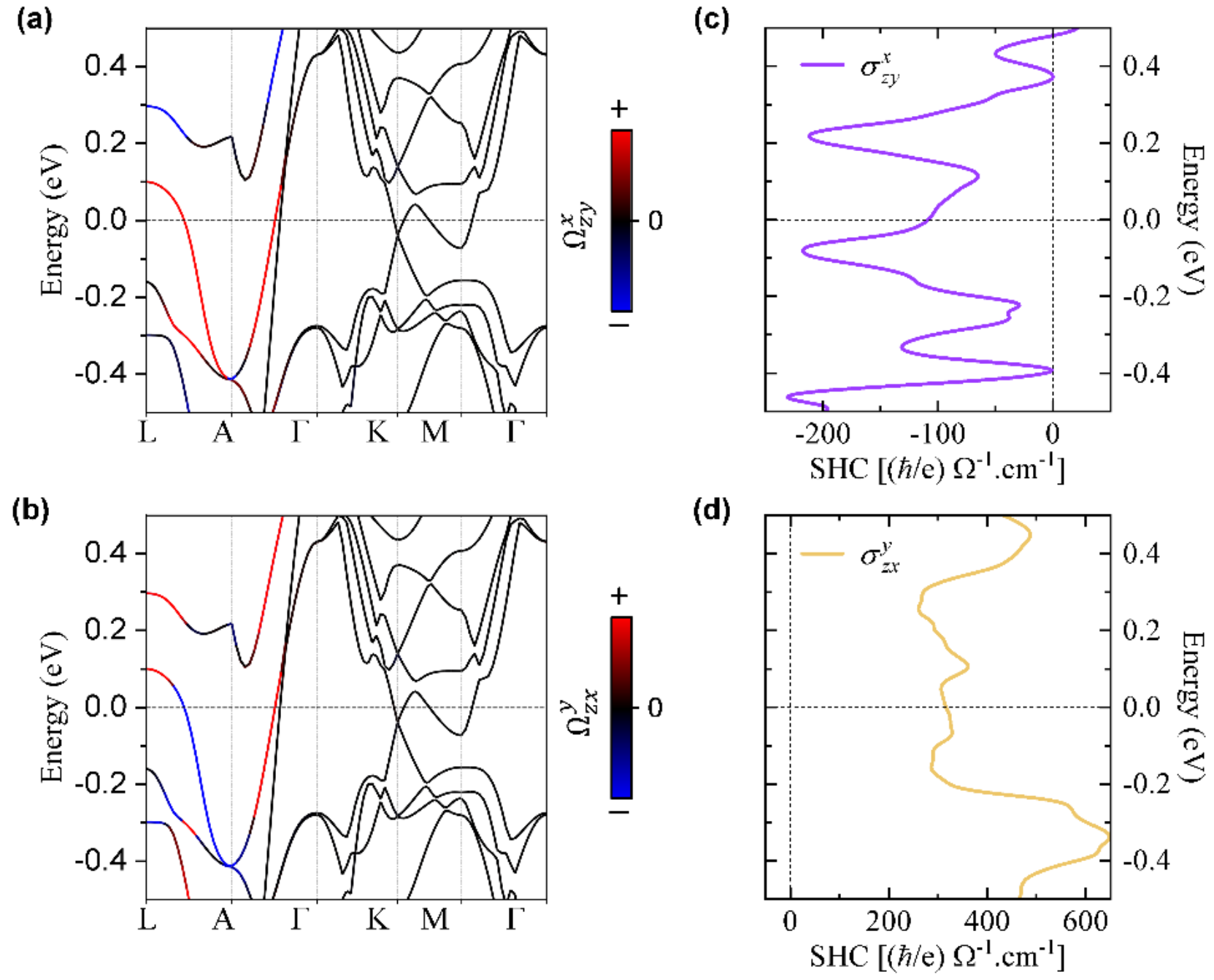

**Figure 5:** ***Ab initio* calculations of electronic band structure and SHC.** Spin Berry curvature-resolved band structure along high-symmetry lines of the Brillouin zone for **a**, $\Omega^x_{zy}$, and **b**, $\Omega^y_{zx}$ of hexagonal noncollinear AF $Mn_3Sn$. Red (blue) denotes positive (negative) contributions. **c**, and **d**, Corresponding energy-dependent $k$-integrated SHC tensor elements accessible experimentally.

We project the corresponding intrinsic spin Berry curvature magnitude, which serves as the microscopic source term of the intrinsic (I)SHE (refer to Eq. 9 and Eq. 10 of the Methods Section 5.5), onto the electronic band structure of $Mn_3Sn$ obtained through DFT, shown in Fig. 5a,b. The corresponding computed SHC tensor elements are shown in Fig. 5c,d. Our findings align with previous *ab initio* studies, revealing that while $\sigma^x_{zy}$ assumes moderate negative values, of order -100 ($\hbar$/e) $\Omega^{-1}$.cm$^{-1}$, at and near the Fermi level, $\sigma^y_{zx}$ assumes significantly larger positive values, of order 300 ($\hbar$/e) $\Omega^{-1}$.cm$^{-1}$. Although the sign of the experimentally reported SHC value of 44.4 ($\hbar$/e) $\Omega^{-1}$.cm$^{-1}$ is consistent with the latter, the magnitude and quasi-isotropic response we observe diverges significantly from the *ab initio* predictions of a sign-changing and crystallographically dependent magnitude of the SHC. Although we did not resolve this discrepancy, we attempt to provide a limited perspective on the potential contributions, in particular extrinsic, responsible for our observations. Furthermore, the electronic band structure of $Mn_3Sn$, even in single-crystal form, remains poorly characterized experimentally [59]. Consequently, the relationship between theoretical models—whether they account for electronic correlations or not [60]—and the actual properties of thin film samples, including their magnetic and electronic configurations, may be overstated.

## 3. Discussions

It is worth noting that experimental reports of SHC or SHA estimates may demonstrate moderate to strong discrepancies with *ab initio* calculations of SHC, as well as among different experimental methodologies. These methodologies include FMR spin-pumping, spin-Seebeck, spin-torque devices, terahertz spintronics emitters, and spin-valves. This phenomenon is not exclusive to $Mn_3Sn$, as indicated in supplementary Table S1; it also occurs in simpler and extensively investigated systems, such as Pt, $\beta$-Ta, and $\beta$-W, all of which are recognized as intrinsically strong SHE materials [43, 53, 61-64]. Even Au, which is intrinsically characterized by low SHC, demonstrates the remarkable sensitivity of SHC—both experimentally and theoretically—to subtle variations in factors such as the microstructure of the material, strain state, Fermi level position, and impurity concentration. This sensitivity is perhaps best illustrated by considering the $\alpha$- and $\beta$-tungsten polymorphs and their corresponding spin-dependent transport properties. In its body-centered cubic ground-state A2-type structure (space group $Im\bar{3}m$, No. 229), the so-called $\alpha$-W demonstrated a negligibly small SHA. However, when tungsten was synthesized as a thin film, its metastable $\beta$-polymorph (of considerably higher electrical resistivity) with topologically close-packed A15-type structure (and space group $Pm\bar{3}n$, No. 223) could be adopted. Notably, this $\beta$-phase displayed substantial SHC (or SHA) values [64], as predicted by first-principles calculations [65]. This example underscores the critical role of crystallographic structure—and, to some extent, the degree of atomic ordering—in shaping the intertwined lattice-, electronic-, and spin-dependent properties of solids, to which the orbital degree of freedom and its recently associated orbital Hall effect can be included [65, 66]. Given these considerations, we contend that a quantitative evaluation of the intrinsic SHC in polycrystalline and amorphous $Mn_3Sn$ films is not feasible. These disordered systems lack the highly oriented, long-range crystalline order necessary for such analysis and may also fail to support the noncollinear inverse triangular AF order responsible for both the intrinsic TR-even and extrinsic TR-odd mechanisms.

Notably, noncollinear AFs, such as $Mn_3Sn$, break both TR and lattice-translational symmetries, as well as the simultaneous TR and inversion symmetry operations. Consequently, even within a linear response framework, responses which are odd under TR symmetry operation are not compelled to vanish, similar to FMs, and in contrast to collinear AFs. Therefore, for $Mn_3Sn$, there exists a second source of non-zero SHC tensor elements, which are listed in Table 1. At the microscopic level, the magnitude of the TR-odd SHC is influenced by the momentum scattering rate. This contrasts with the TR-even SHC, which is of topological origin and is typically calculated under the constant relaxation time approximation.

Consequently, the TR-odd SHC is classified as an extrinsic contribution that can be effectively modulated by an extrinsic disorder source term, irrespective of its specific origin [11]. Specifically, Železný *et al.* [11] demonstrated that, within a straightforward microscopic model, the magnitude of the extrinsic TR-odd SHC elements can be modulated by over an order of magnitude by varying the effective momentum scattering rate (or relaxation time), similarly by over an order of magnitude. Similarly, (I)SHE processes which are odd under time-reversal symmetry (TRS) have been reported in $Mn_3Sn$ single crystals, and popularly coined "*(inverse) magnetic spin Hall effect*" [67], which akin to TRS-breaking ferromagnetically ordered systems are not forced to vanish in noncollinear AF. However, we note that for (0001)-oriented $Mn_3Sn$, none of the non-zero tensor elements of the TR-odd SHC listed in Table 1 are experimentally accessible in our spin-pumping FMR/ISHE setup, and that the magnetic ISHE contributions reported in Ref. [67] are also expected to vanish for the measurement configuration used here.

Lastly, in order to evaluate extrinsic sources of spin-dependent scattering, exquisite spatially sensitive probes which can relate the microstructure of $Mn_3Sn$(0001) films and their magnetic domain structure at the nanoscale are highly desirable, as demonstrated recently by scanning anomalous Nernst effect microscopy [68]. Those extrinsic sources, beyond side-jump scattering on spin-orbit coupled impurities, could involve spin-dependent scattering (or spin depolarization) at grain boundaries, which also act as strong pinning centers for antiferromagnetic domains [68], implying that even for large in-plane magnetic fields, $Mn_3Sn$(0001) films are not single-domain, and that their magnetic anisotropy is distinct from that of $Mn_3Sn$ single crystals.

The continuous spin-pumping process, and resultant DC voltage drop due concomitantly to the ISHE, extrinsic spin-dependent scattering, and to spurious anisotropic magnetoresistance rectification effects complicate the separation of various contributions discussed. To independently evaluate several of the SHC tensor elements listed in Table 1, nonlocal measurements of the (I)SHE in lateral spin-valve structures [69], or in monolithic devices devoid of interfaces that probe the nonlocal transport of charge currents in the diffusive SHE regime [70], may prove beneficial. This approach has previously been employed in $Mn_3Sn$ bulk single crystals [67]. The investigation of single crystals for spin-based interfacial phenomena is often complicated by the inherent challenges associated with preparing pristine interfaces with other spin-active SHE or FM materials. Finally, we suggest that both carefully designed pulsed and time-resolved SOT switching experiments [71, 72], as well as an analysis of the temperature dependence of various measured (I)SHE and AHE signals, conducted either upon lowering the temperature to just above the spin-glass transition temperature (approximately 50 K), or upon approaching the Neél temperature of $Mn_3Sn$ (approximately 420 K), may help to disentangle intrinsic contributions from extrinsic ones. This approach can also shed light on the momentum scattering time-dependent TR-odd term, assuming that the AF order remains essentially unchanged.

## 4. Conclusion

We systematically investigated the thickness-dependent magneto-transport and room-temperature spin-dependent transport properties of heteroepitaxial thin films of the topological Weyl AF $Mn_3Sn$. Crystalline $Mn_3Sn(0001)$ films demonstrated a negligibly small in-plane anomalous Hall conductivity at room temperature, moderately low effective SHA (approximately 0.9%) when accounting for the interfacial spin-transparency, and a quasi-isotropic SHC, which reaches 44.4 ($\hbar$/e) $\Omega^{-1}$.cm$^{-1}$. Additionally, we observed remarkably high values for spin-mixing conductance (28.52 nm$^{-2}$) and interfacial spin-transparency (approximately 72%), surpassing those of commonly investigated spintronic heterostructures [53, 62-65]. These results underscore the potential of $Mn_3Sn$ for applications in SOT technologies. The spin diffusion length estimated in this study, ranging from 15.2 and 25.3 nm, is relatively long and may be influenced by extrinsic factors. This characteristic can be advantageous for the development of planar nonlocal spin-based electronic architectures, such as spin-valves. Furthermore, we observed a negligible exchange bias field at room temperature, indicating that epitaxial $Mn_3Sn$ (0001)/Py heterostructures served as an ideal platform for investigating spin transport at noncollinear antiferromagnet/ferromagnet interfaces.

Conversely, our findings of a quasi-isotropic SHC in otherwise anisotropic epitaxial crystalline $Mn_3Sn$ films underscore the challenges in drawing definitive conclusions regarding the source terms of spin-to-charge conversion. This complexity originated from the phenomenological framework of FMR spin-pumping and ISHE processes within such a complex noncollinear AF hexagonal system. To elucidate the precise microscopic origins—or potentially competing multiple origins—of spin-to-charge interconversion mechanisms in $Mn_3Sn$-based heterostructures, where $Mn_3Sn$ may function as either a spin source, spin sink, or both, careful control experiments are essential. These experiments should include detailed investigations of temperature dependence, thickness dependence, and crystallographic orientation, utilizing a variety of complementary probes. Such probes should evaluate not only the spin and electronic transport properties but also the structural properties, strain states (along with corresponding lattice parameters), and potentially the real electronic band structure of the samples through angle-resolved photoemission spectroscopy.

Our research highlights the potential of crystalline $Mn_3Sn$ films as a moderately efficient spin-to-charge converter at room temperature. To fully leverage the capabilities of $Mn_3Sn$ in next-generation topological AF spintronic devices, future investigations should prioritize the optimization of its spin-dependent properties, including a focused effort on the deliberate engineering of heterointerfaces to enhance performance. Although the spin Hall angle of $Mn_3Sn$ is moderate compared to canonical ISHE materials such as Pt, the observed high spin transparency and high-quality epitaxial interfaces are promising from the antiferromagnetic spintronics standpoint. However, in order to assess the potential of epitaxial $Mn_3Sn$ for *e.g.*, SOT-based magnetic memory applications, the efficiency metrics associated with SOT-switching phenomena must be extracted via other methods such as SOT-FMR and second-harmonic Hall measurements [73], which require micro-patterned $Mn_3Sn$/FM bilayers. The spin-dependent figures of merit reported here can be further enhanced by adjusting the source or sink of the spin current, constructing fully epitaxial heterointerfaces, incorporating spacer layers, and fine-tuning the thickness and composition of the materials. Moreover, to deepen our understanding at a fundamental level, external tuning parameters, such as strain—imposed either epitaxially by the underlying substrate or through piezo- or ferroelectric elements—can provide additional control over both the magnetic order and spin-transport properties, particularly given the piezomagnetic character of AF hexagonal $Mn_3Sn$ [74, 75].

# 5. Methods

## 5.1. Thin films growth

The $Mn_3Sn$ heterostructures were synthesized using magnetron sputtering within a BESTEC ultrahigh vacuum (UHV) system, achieving a base pressure of less than $2 \times 10^{-9}$ mbar and process gas (Ar 5 N) working pressure of $3 \times 10^{-3}$ mbar. The target-to-substrate distance was maintained at 20 cm, and the substrate holder was rotated at 20 rpm during deposition to ensure homogeneous growth. We utilized commercially available single-crystal MgO(111) substrates, with dimensions of 10 × 10 × 0.5 mm, sourced from Crystal GmbH. The Ru buffer layer was initially grown at 500 °C and allowed to cool to room temperature. Subsequently, Mn and Sn were co-sputtered at room temperature from high-purity elemental targets at 56 and 10 W DC power, respectively. The resulting Ru/$Mn_3Sn$ bilayers were heated at a ramp rate of 10 °C $min^{-1}$, followed by post-annealing at 500 °C for 5 min. The Py films were grown *in situ*, from a $Ni_{81}Fe_{19}$ alloyed target, after the sample was cooled to near room temperature (less than 50 °C). All heterostructures were capped *in situ* with a protective 3-nm-thick Si layer, which was deposited through radiofrequency (RF) sputtering from an Si target. This layer naturally oxidizes and passivates when exposed to air.

## 5.2. Structural characterizations

The crystal structure, growth rate, and film thickness were determined through X-ray diffraction and X-ray reflectivity measurements, conducted with a Panalytical X Pert[3] MRD diffractometer utilizing Cu $K_{\alpha 1}$ radiation ($\lambda$ = 1.5406 Å). Surface topography was analyzed using atomic force microscopy from Asylum Research (Oxford Instruments). Crystal structure visualization in Fig. 1a was created in part using VESTA [75].

## 5.3. Magnetometry and magneto-transport measurements

Exchange-bias measurements were performed using a Quantum Design (MPMS3 SQUID-VSM) magnetometer. For magneto-transport measurements, which included assessments of longitudinal and transverse resistivities, unpatterned square-shaped samples were configured in the van der Pauw geometry. Ultrasonic bonding techniques were employed to connect aluminum wires to the corners of the samples. These measurements were performed by utilizing a Quantum Design physical property measurement system (PPMS) with low-frequency alternating current.

## 5.4. Spin-pumping FMR and ISHE measurements

Frequency-dependent FMR measurements were conducted utilizing a NanOsc cryogenic FMR setup integrated with a Quantum Design PPMS system. The samples were positioned in a flip-chip configuration on top of a 200 μm wide coplanar waveguide. Systematic FMR measurements were performed at 300 K across a frequency range of 3–20 GHz. These measurements were executed in an in-plane geometry, maintaining a constant RF field frequency and magnitude (61 μT) while sweeping the external DC magnetic field. To investigate the ISHE, electrical contacts were established on two opposite edges of the sample using silver paste to ensure homogeneous electrical potential. The voltage drops associated with the ISHE during the FMR measurements were recorded using a Keithley 2182A nanovoltmeter. Our spin-pumping FMR/ISHE setup was benchmarked resorting to heterostructures of permalloy/Pt thickness series grown on silicon substrates, for which we find a spin-mixing conductance of 25.9 $nm^{-2}$, a spin Hall angle of 4.5 % and a spin diffusion length of 1.3 nm.

## 5.5. First-principle calculations

We utilized the generalized gradient approximation to describe the exchange-correlation potential, adhering to the Perdew-Burke-Ernzerhof parametrization scheme [77]. A $k$-point grid of 8 × 8 × 8 was utilized, with a total energy convergence criterion set at $10^{-6}$ eV. From DFT calculations, we projected the *ab initio* DFT Bloch wavefunctions onto atomic-orbital-like Wannier functions, resorting to Wannier90 [78]. Furthermore, we generated the corresponding tight-binding (TB) model Hamiltonian, which preserved the full symmetry of the material system under investigation. Within the framework of the (I)SHE, the spin current flowed along the $\alpha$-direction, whereas the generated electric field ($E_\beta$) was oriented along the $\beta$-direction, and spin-polarization ($s$) was aligned along the $\gamma$-direction ($J^{\gamma}_{\alpha}$). These quantities were interconnected through the SHC tensor ($\sigma^{\gamma}_{\alpha\beta}$), expressed as $J^{\gamma}_{\alpha}$=$\sigma^{\gamma}_{\alpha\beta}E_{\beta}$. We redefined the coordinate system ($\alpha$, $\beta$, $\gamma$) to align with the natural coordinate system ($x$, $y$, $z$), constrained by the experimental geometry and out-of-plane [0001] crystallographic direction of our epitaxial $Mn_3Sn$ films. Utilizing the derived TB Hamiltonian, we calculated the intrinsic SHC tensor $\sigma^{\gamma}_{\alpha\beta}$ formulas as follows:

$$\sigma^{\gamma}_{\alpha\beta} = e\hbar \int_{\mathrm{BZ}} \frac{d\boldsymbol{k}}{(2\pi)^3} \sum_{\mathrm{n}} f_{\mathrm{n}\mathbf{k}}\, \Omega^{\hat{S}_{\gamma}}_{\mathrm{n},\alpha\beta}(\mathbf{k}) \,, \tag{9}$$

$$\Omega_{\mathrm{n},\alpha\beta}^{\hat{S}_\gamma}(\mathbf{k}) = -2\mathrm{Im}\sum_{\mathrm{m}\neq\mathrm{n}} \frac{<\mathrm{n}(\mathbf{k})\left|j_\alpha^{\hat{S}_\gamma}\right|\mathrm{m}(\mathbf{k})><\mathrm{m}(\mathbf{k})|\hat{v}_\beta|\mathrm{n}(\mathbf{k})>}{(E_\mathrm{n}(\mathbf{k})-E_\mathrm{m}(\mathbf{k}))^2}, \quad (10)$$

where $j_\alpha^{\hat{S}_\gamma} = \frac{1}{2}\{\hat{v}_\alpha, \hat{S}_\gamma\}$ represents the spin current operator and $\Omega_{\mathrm{n},\alpha\beta}^{\hat{S}_\gamma}(\mathbf{k})$ represents the spin Berry curvature. $\hat{S}_\gamma$ represents the spin operator, $E_\mathrm{n}(\mathbf{k})$ represents the eigenvalue for the $\mathrm{n_{th}}$ eigenstate $|\mathrm{n}(\mathbf{k}) >$ at momentum $\mathbf{k}$, $\hat{v}_{\alpha(\beta)}$ represents the $\alpha(\beta)$ component of the band velocity operator defined by $\hat{v}_{\alpha(\beta)} = \frac{1}{\hbar}\frac{\partial \hat{H}(\boldsymbol{k})}{\partial k_{\alpha(\beta)}}$, and $f_{\mathrm{n}\mathbf{k}}$ represents the Fermi-Dirac distribution function. For the integration in Eq. 10, we employed a uniform $10 \times 10 \times 10$ $k$-grid to facilitate $k$-space summation. Given the experimental setup, the $z$-direction was aligned with the [0001] (or $c$-axis) crystallographic direction of $Mn_3Sn$, whereas the orthogonal $x$- and $y$-directions, which established a Cartesian coordinate system, were oriented along the $[2\bar{1}\bar{1}0]$ and $[01\bar{1}0]$ crystallographic directions of $Mn_3Sn$ (refer to Fig. 1a).

## Acknowledgments

This research was supported by the Horizon 2020 FETPROAC Project No. SKYTOP-824123, “Skyrmion-Topological Insulator and Weyl Semimetal Technology.” C.F. acknowledges financial support by the Deutsche Forschungsgemeinschaft (DFG, German Research Foundation) through the Würzburg-Dresden Cluster of Excellence ctd.qmat – Complexity, Topology and Dynamics in Quantum Matter (EXC 2147, project-id 390858490).

## Authors contributions

C.F. and E.L. proposed and supervised this study. S.N.P. grew the samples, performed the structural characterizations, and performed the magneto-transport and magnetometry experiments under the supervision and advising of E.L. S.N.P. performed the experimental data analysis under the advising of A.M. and E.L. N.M., N.P. and X.F. performed the first-principle calculations and linear transport response theory calculations and analysis under the advising of Y.Z. S.N.P. and E.L. wrote the manuscript, with input from all the authors. S.N.P., N.M. and E.L. prepared the figures. All authors reviewed the manuscript.

## Competing interests

The authors declare no competing interests.

## Corresponding author

Edouard Lesne
Max Planck Institute for Chemical Physics of Solids, 01187 Dresden, Germany
edouard.lesne@cpfs.mpg.de

## Data availability

The data supporting the findings of this study are available from the corresponding author upon reasonable request.

## Additional information

See **Supplementary Information** for: additional structural and magnetic characterizations (Fig. S1), extraction of the resistivity of the $Mn_3Sn$ layer (Fig. S2), FMR dispersion relation and Gilbert damping along orthogonal in-plane directions of $Mn_3Sn$(0001) (Fig. S3); compilation of spin-dependent figures of merit for $Mn_3Sn$ noncollinear AF (Table S1).

# – Supplementary Information –

## Efficient spin-pumping and interfacial spin transport across epitaxial $Mn_3Sn(0001)$ noncollinear antiferromagnet/permalloy interfaces

Surya N. Panda[1], Ning Mao[1], Nikolai Peshcherenko[1], Xiaolong Feng[1], Yang Zhang[2,3], Anastasios Markou[4], Claudia Felser[1], and Edouard Lesne[1,*]

[1]Max Planck Institute for Chemical Physics of Solids, 01187 Dresden, Germany

[2]Department of Physics and Astronomy, University of Tennessee, Knoxville, Tennessee 37996, USA

[3]Min H. Kao Department of Electrical Engineering and Computer Science, University of Tennessee, Knoxville, Tennessee 37996, USA

[4]Physics Department, University of Ioannina, 45110 Ioannina, Greece

* edouard.lesne@cpfs.mpg.de

### S1. Additional structural and magnetic characterizations.

The overall crystalline quality of the $Mn_3Sn$ films was evaluated by measuring the full width at half maximum (FWHM) of rocking curves around the (0002) Bragg diffraction peak. A small 0.67° FWHM [Fig. S1a] confirms minimal mosaicity with high crystalline quality of epitaxial $Mn_3Sn(0001)$ films.

Atomic force microscopy (AFM) imaging is carried out to study the surface topography of $Mn_3Sn$ thin films. In Fig. S1b, an AFM image of a 40 nm-thick $Mn_3Sn$ film is presented, showing an average root-mean-square roughness of about 0.54 nm.

The exchange bias field ($\mu_0 H_{EB}$) was estimated from magnetization curves *M-H* acquired after field cooled (FC) and zero-field cooled (ZFC) procedure. First, the $Mn_3Sn$/Py samples are demagnetized in a large oscillatory in-plane magnetic field at 400 K. The samples are then cooled to the measurement temperature (300 K or 5 K) in either a 1T in-plane magnetic field (FC) or in ZFC condition, where an $M(\mu_0 H)$ hysteresis loop is acquired and from which we evaluate the magnitude of $\mu_0 H_{EB}$ and of the coercive field ($\mu_0 H_c$). Fig. S1c display the ZFC and FC hysteresis loops at 5 K for a 40 nm-thick $Mn_3Sn$/Py(12 nm) heterostructure, where $\mu_0 H_{EB}$ and $\mu_0 H_c$ are found to be 20 mT and 4.3 mT, respectively. However, at room temperature (as shown in Fig. 1f of the manuscript), exchange bias is not observed, while $\mu_0 H_c$ decreases significantly to only 1.3 mT.

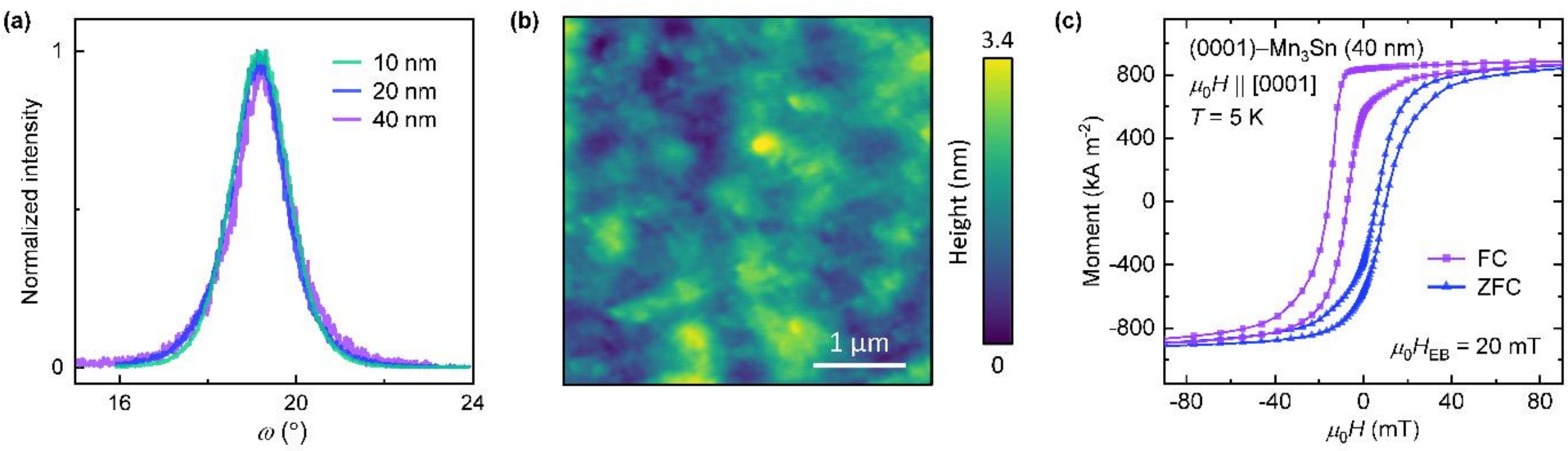

**Figure S1 – Structural and magnetic characterizations a**, Rocking curve ($\omega$-scan) of (0002) Bragg diffraction peaks for $Mn_3Sn$(0001) films of various thicknesses grown on MgO(111). **b**, AFM topography image of a $Mn_3Sn$ (40 nm) sample. **c**, Hysteresis loop of the $Mn_3Sn$ (40 nm)/Py (12 nm) heterostructure measured at 5 K for field cooled (FC) and zero-field cooled (ZFC) measurement protocols.

## S2. Extraction of the electrical resistivity of $Mn_3Sn$.

For longitudinal resistivity measurements, we employed a Physical Properties Measurement System (PPMS) Quantum Design cryostat and ultrasonically bonded aluminum wires to the corners of square-shaped samples in a van der Pauw geometry. We employ low frequency current excitation and lock-in amplifier detection of the voltage drop. Fig. S2a shows the total longitudinal conductivity of bilayer Ru/$M_3Sn$ samples, with $Mn_3Sn$ thickness varying from 20 nm to 80 nm and the Ru buffer layer thickness kept fixed at 3 nm.

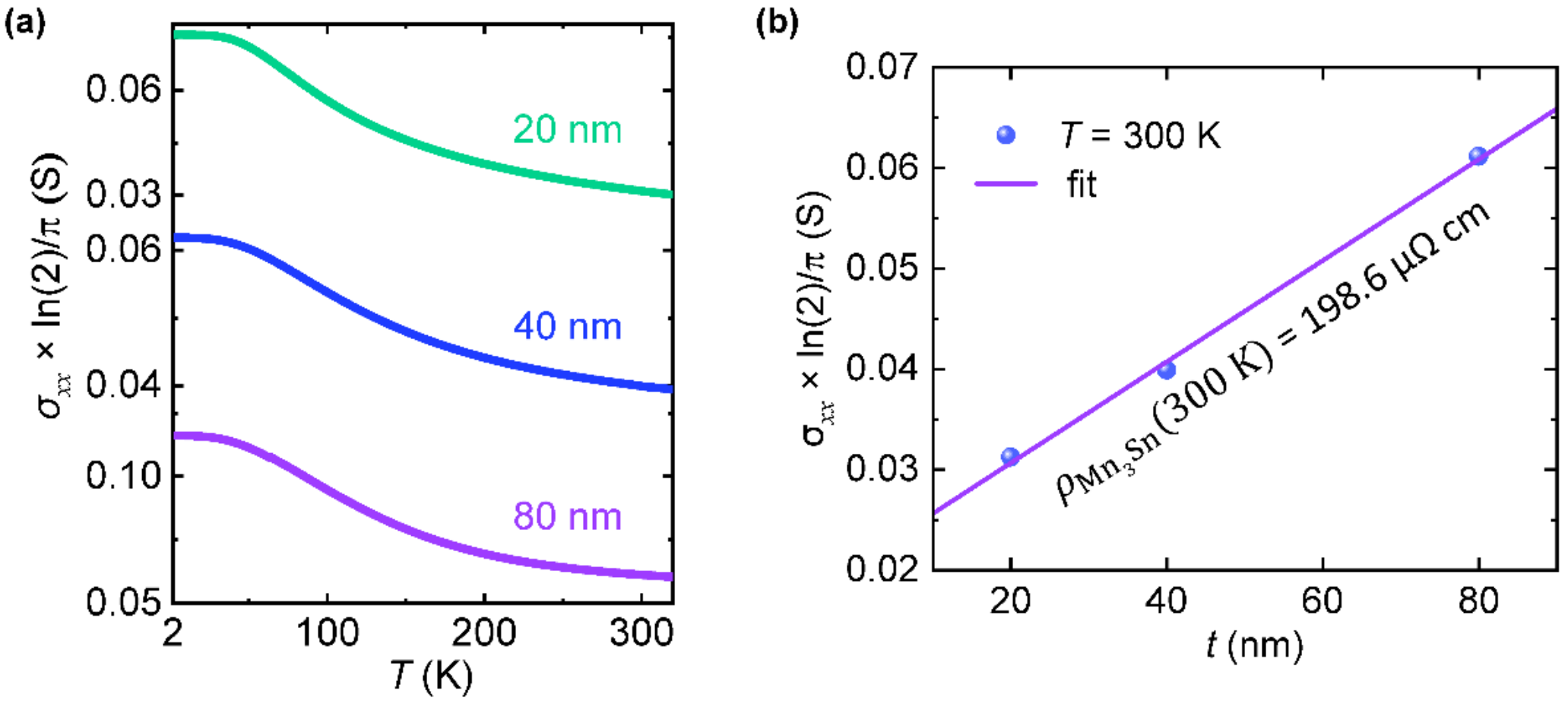

**Figure S2 – Thickness and *T*-dependent electrical transport in $Mn_3Sn$ samples. a**, Temperature-dependence of longitudinal conductivity ($\sigma_{xx}$) for (111)–MgO//Ru(3 nm)/$Mn_3Sn$($t$ = 20, 40, 80 nm) samples. **b**, $Mn_3Sn$ thickness-dependent longitudinal conductivity at 300 K. Solid line represents a linear fit [Suppl. Eq. (2)] which allows the extraction of the resistivity of our $Mn_3Sn$ films.

In turn we model the charge transport in our heterostructure using a parallel resistor model composed of two resistors ($R_{\mathrm{Mn_3Sn}}$ and $R_{\mathrm{Ru}}$), where the total measured resistance of the bilayer is $R_{xx}$, such as:

$$\frac{1}{R_{xx}} = \frac{1}{R_{\mathrm{Mn_3Sn}}} + \frac{1}{R_{\mathrm{Ru}}} \ . \tag{1}$$

It follows that the sheet conductance $\sigma_s$ of the bilayer can be expressed as:

$$\sigma_s = \sigma_{xx} \times \frac{\ln(2)}{\pi} = \frac{t}{\rho_{\mathrm{Mn_3Sn}}} + \frac{d}{\rho_{\mathrm{Ru}}} , \tag{2}$$

where $\frac{\ln(2)}{\pi}$ is the van der Pauw factor, the resistivities of the $Mn_3Sn$ and Ru layers are denoted by $\rho_{\mathrm{Mn_3Sn}}$ and $\rho_{\mathrm{Ru}}$, $t$ and $d$ are the thicknesses of the $Mn_3Sn$ and Ru layers, respectively. By fitting the $Mn_3Sn$ thickness-dependent $\sigma_{xx}$ (Fig. S2b) using Suppl. Eq. (2), we have extracted the values for $\rho_{\mathrm{Mn_3Sn}}$ and $\rho_{\mathrm{Ru}}$. Thus, at 300 K, $\rho_{\mathrm{Mn_3Sn}}$ = 198.6 μΩ.cm and $\rho_{\mathrm{Ru}}$ =14.6 μΩ.cm.

## S3. FMR dispersion relation and Gilbert damping along orthogonal in-plane directions of $Mn_3Sn(0001)$.

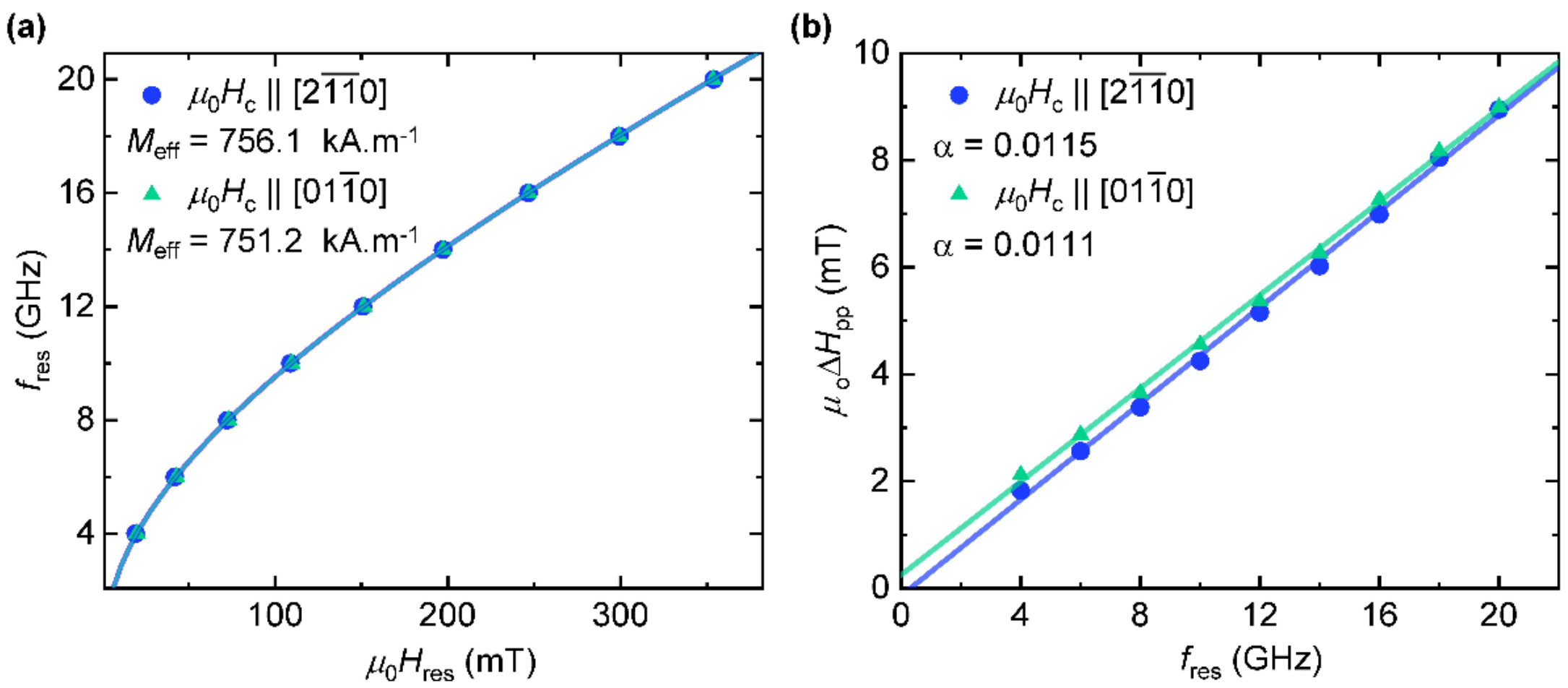

**Figure S3 – Crystallographically-independent FMR response of 40 nm-thick $Mn_3Sn(0001)$/Py(12 nm) heterostructure. a,** Kittel dispersion relation $f_{res}(H_{res})$, and corresponding fits (Eq. 2, solid lines), and **b,** peak-to-peak FMR linewidth and correspondingly evaluated effective Gilbert-damping α (solid lines are fits to Eq. 3) for the DC magnetic field applied along two orthogonal in-plane directions $[2\bar{1}\bar{1}0]$ and $[01\bar{1}0]$ of $Mn_3Sn(0001)$.

**Table S1 – Compilation of spin-dependent figures of merit reported for $Mn_3Sn$-based systems.** ST-FMR: spin-torque ferromagnetic resonance; TR-MOKE: time-resolved magneto-optical Kerr effect. Ref. S2 quotes a spin coherence length, which we assimilate to the spin diffusion length here.

| [Ref.] | $Mn_3Sn$ ordering | $Mn_3Sn$ form | Thickness range (nm) | Spin injector | Spin injection method | Detection method | Spin Hall angle (%) | Spin Hall conductivity [$(\hbar/e)\ \Omega^{-1}\ cm^{-1}$] | Spin-mixing conductance ($nm^{-2}$) | Spin diffusion length (nm) | Interfacial spin transparency (%) | Damping-like/field-like torque ratio | SOT efficiency |
|---|---|---|---|---|---|---|---|---|---|---|---|---|---|
| **This work** | Epitaxial | Thin film | 10–80 | Py | Spin-pumping | ISHE | 0.88 | 44.4 | 28.52 | 15.2–25.3 | 72 | | |
| **[S1]** | Poly-crystalline | Thin film | 10 | YIG | Spin-pumping & spin Seebeck | ISHE | 18.1 | | | | | | |
| **[S2]** | Epitaxial | Thin film | 5–45 | Pt | Circularly polarized light | TR-MOKE | | | | ≈ 15 | | | |
| **[S3]** | Single-crystal | Bulk | | Py | ST-FMR | ST-FMR/ISHE | | 1400 | | | | | 0.5 |
| **[S4]** | Epitaxial | Thin film | 5-20 | Py | ST-FMR | ST-FMR/ISHE | | | | | | | 0.06 |
| **[S5]** | Epitaxial | Thin film | 30 | Py | ST-FMR | ST-FMR/ISHE & MOKE | 5.3 | | | | | 0.5 | |
| **[S6]** | Epitaxial | Thin film | 3–12 | Py | ST-FMR | ST-FMR/ISHE | | | | | | 20 | |
| **[S7]** | Poly-crystalline | Thin film | 50–70 | Py | Electrical spin injection | Nonlocal ISHE | 5.3 | 46.99 | | 0.75 | | | |
| **[S8]** | Poly-crystalline | Thin film | 1–10 | Py | ST-FMR | ST-FMR/ISHE | | -702 | 15.4 | | | | |

## Supplementary References